\documentclass{ws-mpla}

\begin{document}

\markboth{S.Meljanac et al.}
{Kappa-deformed Snyder spacetime}

\def\p{\partial}
\def\x{\hat x}

\catchline{}{}{}{}{}

\title{KAPPA-DEFORMED SNYDER SPACETIME}

\author{\footnotesize S.MELJANAC\footnote{meljanac@irb.hr},
\footnotesize D.MELJANAC\footnote{dmeljan@irb.hr},
\footnotesize A.SAMSAROV\footnote{asamsarov@irb.hr},
\footnotesize M.STOJIC\footnote{marko.stojic@zg.htnet.hr}}

\address{Rudjer Boskovic Institute,\\ Bijenicka  c.54, HR-10002 Zagreb,
Croatia}

\maketitle


\begin{abstract}
 We present Lie-algebraic deformations of Minkowski space with
       undeformed Poincar\'{e} algebra. These deformations interpolate
       between Snyder and $\kappa$-Minkowski space. We find
       realizations of noncommutative coordinates in terms of
       commutative coordinates and derivatives. Deformed Leibniz rule, the coproduct structure and star
      product are found. Special cases, particularly Snyder and
       $\kappa$-Minkowski in Maggiore-type realizations are discussed.
       Our construction leads to a new class of deformed special
       relativity theories.

\end{abstract}

\ccode{PACS Nos.: 03.65.Fd, 11.10.Nx, 11.30.Cp.}

\section{Introduction}    

 Noncommutative (NC) physics has became an integral part of
    present-day high energy physics theories. It reflects a structure
    of space-time which is modified in comparison to space-time
    structure underlying the ordinary commutative physics. This
    modification of space-time structure is a natural consequence of
    the appearance of a new fundamental length scale known as Planck
    length \cite{Doplicher:1994zv,Doplicher:1994tu}.
 There are two main physical contexts within which a signal
    for the existence of a Planck length scale appears.
     The first one lies within a loop quantum gravity framework in
    which the Planck length plays a fundamental role. There, a
    presence of a new
    fundamental length scale leads after quantization to discrete
    eigenvalues of
    the area and volume operators. It appears that in loop quantum
    gravity, the area and volume operators have
    discrete spectra, with minimal eigenvalue proportional to a square
    and cube of the Planck length, respectively. The second physical
    context where one can find a signal for the existence of a
    fundamental length scale comes
     from some observations of ultra-high energy cosmic rays
    which seem to contradict the usual understanding of some
    astrophysical processes like, for example, electron-positron
    production in collisions of high energy photons. It turns out that
    deviations observed in these processes can be explained by
    modifying dispersion relation in such a way as to incorporate the
    fundamental length scale \cite{AmelinoCamelia:2000zs}. NC space-time has also been revived in
    the paper by Seiberg and Witten \cite{Seiberg:1999vs} where NC manifold emerged in a
    certain low energy limit of open strings moving in the background
    of a two form gauge field.

   As a new fundamental, observer-independent quantity, Planck length
   is incorporated in kinematical theory within the framework of the so
   called doubly special relativity theory (DSR)
 \cite{AmelinoCamelia:2000mn,AmelinoCamelia:2000ge}. The idea
   that lies behind DSR is that there exist two observer-independent
   scales, one of velocity, identified with the speed of light, and
   the other of mass, which is expected to be of the order of Planck mass. 
   It can also be considered as a semi-classical, flat space limit
   of quantum gravity in a similar way special relativity is a limit
   of general relativity and, as such, reveals a structure of
   space-time when the gravitational field is switched off.

    Following the same line of reasoning, the symmetry algebra for
    doubly special relativity can be obtained by deforming the
    ordinary Poincare algebra to get some kind of a quantum (Hopf)
    algebra, known as $\kappa$-Poincar\'{e} algebra
 \cite{Lukierski:1991pn,Lukierski:1993wx}, so that 
   $\kappa$-Poincar\'{e} algebra is in the same relation to DSR theory
    as the standard  Poincar\'{e} algebra is related to special
    relativity.

 $\kappa$-Poincar\'{e} algebra is an algebra that describes in a
 direct way only the energy-momentun sector of the DSR
 theory. Although this sector alone is insufficient to set up physical
 theory, the Hopf algebra structure makes it possible to extend the
 energy-(angular)momentum algebra to the algebra of space-time.
 It is shown in \cite{KowalskiGlikman:2002we}  
 that different representations (bases) of $\kappa$-Poincar\'{e}
 algebra correspond to different DSR theory. However, the resulting
 space-time algebra, obtained by the extension of energy-momentum
 sector, is independent of the representation, i.e. energy-momentum
 algebra chosen \cite{KowalskiGlikman:2002we,KowalskiGlikman:2002jr}.

 It is also shown in \cite{KowalskiGlikman:2002jr} that there exists a
 transformation which maps $\kappa$-Minkowski space-time into
 space-time with noncommutative structure described by the algebra
 first introduced by Snyder \cite{snyder}. In
 \cite{KowalskiGlikman:2002jr}, the use of Snyder algebra
 provided NC space-time structure of Minkowski space with undeformed
 Lorentz symmetry. In the same paper it is argued that the algebra
 introduced by Snyder provides a structure of configuration space for
  DSR and thus can be used to construct the second order  particle
 Lagrangian, what would make it possible to define physical
 four-momenta determined by the particle dynamics. This would be
 significant step forward because the theoretical development in this
 area has been mainly kinematical so far. One such dynamical picture
 has been given recently in \cite{Ghosh:2006cb,Ghosh1} where it was shown that
 reparametrisation symmetry of the proposed Lagrangian, together with
the appropriate change of variables and conveniently
 chosen gauge fixing conditions, leads to an algebra which is a
 combination of $\kappa$-Minkowski and Snyder algebra. This
 generalized type of algebra describing noncommutative structure of
 Minkowski space-time is shown to be consistent with the
 Magueijo-Smolin dispersion relation. This type of NC space is also
 considered in \cite{Chatterjee:2008bp}. It has to be stressed that NC spaces
 in neither of these papers are of Lie-algebra type.

   In order to fill this gap, in the present paper we unify
  $\kappa$-Minkowski and Snyder space in a more general NC space
  which is of a Lie-algebra type, with Lorentz generators and NC
  space-time coordinates closing the Lie algebra.
In addition, it is 
    characterized by the
   undeformed Poincar\'{e} algebra and deformed coalgebra.
   In other words, we shall consider a type of NC space which
  interpolates between
 $\kappa$-Minkowski space and Snyder space in a Lie-algebraic way and has
    all deformations contained in the coalgebraic
  sector. Particularly, in this paper we shall be interested in
  finding a coproduct for translation generators, which corresponds to
 a generalized momentum addition rule.
  First example of NC space with undeformed Poincar\'{e} algebra, but with deformed coalgebra
    is given by Snyder \cite{snyder}. Some other types of NC spaces are
   also considered within the approach in which the Poincar\'{e}
   algebra is undeformed and coalgebra deformed, in particular the
   type of NC space with $\kappa$-deformation \cite{KowalskiGlikman:2002we,KowalskiGlikman:2002jr,Meljanac:2006ui,KresicJuric:2007nh,Meljanac:2007xb}.
  Here we present a broad class of Lie-algebra type deformations with the same
   property of having deformed coalgebra, but undeformed algebra.
    The investigations carried out in this paper  are along the track
  of developing general techiques of calculations,
              applicable for a widest possible class of NC spaces and
 as such  are a continuation of the work done in a series of previous
   papers 
 \cite{Meljanac:2006ui,KresicJuric:2007nh,Meljanac:2007xb,Jonke:2002kb,Jonke1,Durov:2006iv,Meljanac:2008ud,Meljanac:2008pn}.
   The methods used in these investigations were taken over from the
  Fock space analysis carried out in \cite{Doresic:1994zz,palmil1,palmil2,bardek,bardek1,bardek2}.

   The plan of the paper is as follows. In section 2 we introduce a type
   of deformations of Minkowski space that have a structure of a Lie
   algebra and which interpolate between $\kappa$-type of deformations and 
   deformations of the Snyder type.
      In section 3 we  analyze realizations of NC space in terms
   of operators belonging to the undeformed Heisenberg-Weyl algebra.
 Section 4 is devoted to an analysis of the effects which 
   deformations we are considering have on the coalgebraic structure of the symmetry algebra. 
   In the same section we
   specialize the general results obtained to some interesting special
   cases, such as Snyder space and $\kappa$-Minkowski space.

\section{Noncommutative coordinates  and Poincar\'{e} algebra}

We are considering a Lie algebra type of noncommutative (NC) space generated by
the coordinates $\x_0, \x_1,\ldots ,\x_{n-1},$ satisfying the commutation
relations
\begin{equation} \label{2.1}
[\x_{\mu},\x_{\nu}]=i(a_{\mu}\x_{\nu}-a_{\nu}\x_{\mu})+s M_{\mu
\nu},
\end{equation}
where indices $\mu,\nu=0,1\dots,n-1$ and $a_0,a_1,\dots ,a_{n-1}$ are componenets
of a four-vector $a$ in Minkowski space whose metric signature is 
$~ \eta_{\mu\nu} = diag(-1,1,\cdot \cdot \cdot, 1).$ The quantities
$a_{\mu}$ and $s$ are deformation parameters which measure a degree of
deviation from standard commutativity. They are related to length
scale characteristic for distances where it is supposed that noncommutative character of
space-time becomes important. When parameter $s$ is set to zero,
noncommutativity (\ref{2.1}) reduces to covariant version of
$\kappa$-deformation, while in the case that all components of a
four-vector $a$ are set to $0,$
 we get the type of NC space considered
for the first time by Snyder \cite{snyder}. The NC space of this type
has been annalyzed in the literature from different points of view \cite{Battisti:2008xy,Guo:2008qp,Romero:2004er,Banerjee:2006wf,Glinka:2008tr}.

 The symmetry of
the deformed space (\ref{2.1}) is assumed to be described by an undeformed Poincar\'{e}
algebra, which is generated by  
 generators $M_{\mu\nu}$ of the  Lorentz algebra and generators
 $D_{\mu}$ of translations. This means that generators $M_{\mu\nu},~ M_{\mu \nu} = -M_{\nu \mu}, $
 satisfy the standard, undeformed commutation relations,
\begin{equation} \label{2.2a}
[M_{\mu\nu},M_{\lambda\rho}] =
\eta_{\nu\lambda}M_{\mu\rho}-\eta_{\mu\lambda}M_{\nu\rho}
-\eta_{\nu\rho}M_{\mu\lambda}+\eta_{\mu\rho}M_{\nu\lambda}, 
\end{equation}
with the identical statement as well being true for the generators $D_{\mu},$
\begin{equation} \label{2.5}
[D_{\mu},D_{\nu}] =0, 
\end{equation}
\begin{equation}  \label{2.5a}
[M_{\mu\nu},D_{\lambda}] = \eta_{\nu\lambda}
D_{\mu}-\eta_{\mu\lambda}  D_{\nu}. 
\end{equation}
The undeformed Poincar\'{e} algebra, Eqs.(\ref{2.2a}),(\ref{2.5}) and (\ref{2.5a})
define the energy-momentum sector of the theory considered. However, full
description requires space-time sector as well. Thus, it is of
interest to extend the algebra (\ref{2.2a}),(\ref{2.5}) and (\ref{2.5a})
so as to include NC coordinates $\x_0, \x_1,\ldots ,\x_{n-1},$
and to consider the action of Poincar\'{e} generators on NC
space (\ref{2.1}),
\begin{equation} \label{2.3}
[M_{\mu\nu},\x_{\lambda}]=\x_{\mu}\, \eta_{\nu\lambda}-\x_{\nu}\,
\eta_{\mu\lambda}-i\left(a_{\mu}\, M_{\nu\lambda}-a_{\nu}\,
M_{\mu\lambda}\right).
\end{equation}
The main point is that commutation relations (\ref{2.1}),(\ref{2.2a})
 and (\ref{2.3}) define a Lie algebra generated by Lorentz generators 
$M_{\mu\nu}$ and $\x_{\lambda}.$
 The necessary and sufficient condition for consistency of an
extended algebra, which includes generators $M_{\mu\nu}, ~D_{\mu}$ and $\x_{\lambda},$
 is that Jacobi identity holds for all
combinations of the generators $M_{\mu\nu},$ $D_{\mu}$ and $\x_{\lambda}.$
Particularly, the algebra generated by $D_{\mu}$ and $\x_{\nu}$ is a deformed
Heisenberg-Weyl algebra and we require that this algebra has to be of the form,
\begin{equation} \label{2.6}
[D_{\mu},\x_{\nu}] = \Phi_{\mu\nu}(D),
\end{equation}
where $ \Phi_{\mu\nu}(D)$ are some functions of generators $D_{\mu},$
which are required to satisfy the boundary condition
$ \Phi_{\mu\nu}(0)=\eta_{\mu\nu}.$ This condition means that deformed NC
 space, together with the corresponding coordinates, reduces to
 ordinary commutative space in the limiting case of vanishing
 deformation parameters, $a_{\mu}, s \rightarrow 0.$ 

One certain type of noncommutativity, which interpolates between
Snyder space and $\kappa$-Minkowski space, has already been investigated
 in  \cite{Ghosh:2006cb,Ghosh1,Chatterjee:2008bp} in the context of
 Lagrangian particle dynamics. However, in these papers algebra
 generated by NC coordinates and Lorentz generators is not
 linear and is not closed in the generators of the algebra. Thus, it is not of 
Lie-algebra type. In contrast to this, here we consider an algebra
 (\ref{2.1}),(\ref{2.2a}),(\ref{2.3}), which is of
  Lie-algebra type, that is, an algebra
 which is linear in all generators and
 Jacobi identities are satisfied for all combinations of generators of
 the algebra. Besides that, it is important to note that, once having
 relations (\ref{2.1}) and (\ref{2.2a}), there exists only one possible
 choice for the commutation relation between $M_{\mu\nu}$ and $\x_{\lambda},$
which is consistent with Jacobi identities and makes Lie algebra to close, 
 and this unique choice is given by the commutation relation (\ref{2.3}).

In subsequent considerations we shall be interested in 
possible realizations of the space-time algebra (\ref{2.1}) in terms
of canonical commutative space-time coordinates $X_{\mu},$
\begin{equation} \label{2.9}
[X_{\mu},X_{\nu}]=0,
\end{equation}
which, in addition, with derivatives
 $D_{\mu} $ close the undeformed Heisenberg algebra,
\begin{equation} \label{2.10}
[D_{\mu},X_{\nu}] = \eta_{\mu\nu}.
\end{equation}
From the beginning, the generators $D_{\mu} $ are considered as
deformed derivatives conjugated to $\x$ through the commutation
relation (\ref{2.6}). However,  
in the whole paper we restrict ourselves to natural choice \cite{Meljanac:2007xb} in which
deformed derivatives are identified with the ordinary derivatives, 
$D_{\mu} \equiv \frac{\partial}{\partial X^{\mu}}$.

Thus, our aim is  to find a class of covariant $\Phi_{\alpha\mu}(D)$
realizations,
\begin{equation} \label{2.7}
\x_{\mu} =X^\alpha \Phi_{\alpha\mu}(D),
\end{equation}
satisfying the boundary conditions
$~ \Phi_{\alpha\mu}(0)=\eta_{\alpha\mu},~$ and
 commutation relations (\ref{2.1}) and (\ref{2.3}).
In order to complete this task, we introduce the standard coordinate representation
of the Lorentz generators $M_{\mu\nu},$
\begin{equation} \label{2.8}
M_{\mu\nu} = X_{\mu}D_{\nu}- X_{\nu}D_{\mu}.
\end{equation}
All other commutation relations, defining the extended algebra, are then automatically satisfied, as
well as all Jacobi identities among $\x_\mu$, $M_{\mu\nu},$ and
$D_{\mu}.$ This is assured by the construction (\ref{2.7}) and (\ref{2.8}).
In the next section we turn to problem of finding an explicit $\Phi_{\mu\nu}(D)$ realizations
(\ref{2.7}).

\section{Realizations of NC coordinates}

Let us define general covariant realizations:
\begin{equation} \label{3.1}
\x_{\mu} = X_{\mu}\varphi + i(aX)\left(D_{\mu}\,
\beta_1+ia_{\mu}D^2\, \beta_2\right)+i(XD)
\left(a_{\mu}\gamma_1+i(a^2 - s) D_{\mu}\, \gamma_2\right),
\end{equation}
where $\varphi$, $\beta_i$ and $\gamma_i$ are functions of
$A=ia_{\alpha}D^{\alpha}$ and $B=(a^2-s)D_{\alpha}D^{\alpha}$. We further
impose the boundary condition that $\varphi(0,0)=1$ as the parameters of
 deformation $a_{\mu} \rightarrow 0$ and
 $s \rightarrow 0.$ In this way we assure that $\x_{\mu}$
reduce to ordinary commutative coordinates in the limit of vanishing deformation.

It can be shown that Eq.(\ref{2.3}) requires the following set of
equations to be satisfied,
$$
\frac{\p\varphi}{\p A}=-1,\qquad \frac{\p\gamma_2}{\p A}=0, \qquad
\beta_1=1, \qquad \beta_2=0, \qquad \gamma_1=0. $$
Besides that, the commutation relation (\ref{2.1}) leads to the additional two equations,
\begin{equation}
\varphi(\frac{\p\varphi}{\p A}+1)=0,
\end{equation}
\begin{equation}
(a^2-s)[2(\varphi+A)\frac{\p\varphi}{\p
B}-\gamma_2(A\frac{\p\varphi}{\p A}+2B\frac{\p\varphi}{\p
B})+\gamma_2 \varphi]-a^2\frac{\p\varphi}{\p A}-s=0.
\end{equation}
The important result of this paper is that
all above required conditions are solved by a general form of
realization which in a compact form can be written as
\begin{equation} \label{3.2}
\x_{\mu}=X_{\mu}(-A+f(B))+i(aX)D_{\mu}-(a^2-s)(XD)D_\mu\gamma_2,
\end{equation}
where $\gamma_{2}$ is necessarily  restricted to be
\begin{equation} \label{3.3}
\gamma_2=-\frac{1+2f(B)\frac{d f(B)}{d B}}{f(B)-2B\frac{d
f(B)}{d B}}.
\end{equation}
From the above relation we see that $\gamma_{2}$ 
 is not an independent function, but instead is related 
to generally an arbitrary function $ f(B)$, which has to satisfy the boundary condition $f(0)=1$.
Also, it is readily seen from the realization (\ref{3.2}) that
 $~\varphi ~$ in (\ref{3.1}) is given by $~\varphi = -A + f(B).$ 
Various choices of the function $f(B)$ lead to different realizations
of NC space-time algebra (\ref{2.1}).
The particularly interesting cases are
for $f(B)=1$  and  $ f(B)=\sqrt{1-B}$.

It is now straightforward to show that the deformed Heisenberg-Weyl
algebra (\ref{2.6}) takes the form  
\begin{equation} \label{3.4}
[D_{\mu},\x_{\nu}]=\eta_{\mu\nu}(-A+f(B)) +i a_\mu
D_\nu+(a^2-s)D_\mu D_\nu \gamma_2
\end{equation}
and that the Lorentz generators $M_{\mu\nu}$ can be expressed in terms of
NC coordinates as
\begin{equation} \label{2.4}
M_{\mu\nu}=(\hat{x}_{\mu}D_{\nu}-\hat{x}_{\nu}D_{\mu})\frac{1}{\varphi}.
\end{equation}
We also point out that in the special case when realization of NC space (\ref{2.1}) 
is characterized by the function $ f(B)=\sqrt{1-B}$, there exists a
unique element $ Z$
satisfying:
\begin{equation} \label{2.13}
[Z^{-1},\x_{\mu}] = -ia_{\mu} Z^{-1}+sD_{\mu}, \qquad [Z,D_{\mu}] =0.
\end{equation}
From these two equations it follows
\begin{equation} \label{2.15}
 [Z,\x_{\mu}] = ia_{\mu} Z-sD_{\mu}Z^2, \qquad \x_{\mu}Z\x_{\nu} =\x_{\nu}Z\x_{\mu}.
\end{equation}
The element $Z$ is  a generalized shift operator \cite{KresicJuric:2007nh} and its expression
in terms of $A$ and $B$ can be shown to have the form
\begin{equation} \label{16}
  Z^{-1} = -A + \sqrt{1 - B}.
\end{equation}
As a consequence, the Lorentz generators can be expressed in terms of
$Z$ as
\begin{equation} \label{2.17}
M_{\mu\nu}=(\x_{\mu}D_{\nu}-\x_{\nu}D_{\mu})Z,
\end{equation}
and one can also show that the relation
\begin{equation} \label{2.18}
[Z^{-1},M_{\mu\nu}] = -i(a_{\mu}D_{\nu}-a_{\nu}D_{\mu})
\end{equation}
holds.
In the rest of paper we shall only be interested  in the realizations
 determined by $ f(B)=\sqrt{1-B}$.

\section{Leibniz rule and coalgebra}

  The symmetry
 underlying deformed Minkowski space, characterized by the commutation relations
   (\ref{2.1}), is the deformed Poincar\'{e} symmetry which can most
 conveniently be described in terms of quantum Hopf algebra.
 As was seen in relations (\ref{2.2a}),(\ref{2.5}) and (\ref{2.5a}),
 the algebraic sector of this deformed symmetry is the same as that of undeformed 
Poincar\'{e} algebra. However, the action of Poincar\'{e} generators
 on the deformed Minkowski space is deformed,
  so that the whole deformation
 is contained in the coalgebraic sector. This means that the Leibniz
 rules, which describe the action of $M_{\mu\nu}$ and $D_{\mu}$ generators,
will no more have the standard form, but instead will be deformed
and will depend on a given $\Phi_{\mu\nu}$ realization.

  Generally we find that in a given $\Phi_{\mu\nu}$ realization we can
  write \cite{KresicJuric:2007nh,Meljanac:2007xb}
\begin{equation} \label{4.1}
 e^{ik\x} |0> ~ = ~ e^{iK_{\mu}(k)X^{\mu}}
\end{equation}
and
\begin{equation} \label{4.2}
 e^{ik\x} (e^{iqX}) ~ = ~ e^{iP_{\mu}(k,q)X^{\mu}},
\end{equation}
where $k\x = k^{\alpha}X^{\beta}\Phi_{\beta\alpha} (D). $
In (\ref{4.1})
 we have introduced the vacuum state $|0> ~ \equiv 1$ 
as a unit element in the universal
 enveloping algebra understood as a module over the deformed Weyl
 algebra, which is generated by
  $\x_{\mu}$ and $D_{\mu}, ~~ \mu = 0,1,...,n-1,$ and allows for 
  infinite series in $D_{\mu}$. This vacuum state is defined by
\begin{eqnarray}  \label{3.9}
\phi(X)|0> & \equiv & \phi(X) \cdot 1 ~ = ~ \phi(X),\\
D_{\mu} |0> & \equiv & D_{\mu} 1 ~ = ~0, \qquad  M_{\mu\nu}|0> ~ = ~ 0  \label{3.10},
\end{eqnarray}
with $\phi(X)$ belonging to a space  of ordinary functions in commutative
coordinates.
 It is also understood that
NC coordinates  $\x,$ appearing in (\ref{4.1}) and (\ref{4.2})  refer
 to some particular realization (\ref{3.2}),
i.e. they are assumed to be represented by (\ref{3.2}).

The quantities $K_{\mu}(k)$ are readily identified as $K_{\mu}(k) = P_{\mu}(k,0)$
and $P_{\mu}(k,q)$ can be found by calculating the expression
\begin{equation} \label{4.3}
  P_{\mu}(k,-iD) ~ = ~ e^{-ik\x} (-iD_{\mu}) e^{ik\x},
\end{equation}
where it is assumed that at the end of calculation the identification
$q=-iD$ has to be made.
One way to explicitly evaluate the above expression is by using the
BCH expansion perturbatively, order by order. To avoid this tedious
procedure, we can turn to much more elegant method to obtain the
quantity $P_{\mu}(k,-iD)$. This consists in writing the differential equation
\begin{equation} \label{4.4}
 \frac{dP_{\mu}^{(t)}(k,-iD)}{dt} ~ = ~
 \Phi_{\mu\alpha}(iP^{(t)}(k,-iD)) k^{\alpha},
\end{equation}
satisfied by the family of operators $P_{\mu}^{(t)}(k,-iD),$ defined as
\begin{equation} \label{4.5}
  P_{\mu}^{(t)}(k,-iD) ~ = ~ e^{-itk\x} (-iD_{\mu}) e^{itk\x}, \qquad 0\leq t\leq 1,
\end{equation}
and parametrized with the free parameter $t$ which belongs to the interval
$0\leq t\leq 1.$ The family of operators (\ref{4.5}) represents the
generalization of the quantity $P_{\mu}(k,-iD),$ determined by (\ref{4.3}),
namely, $P_{\mu}(k,-iD) = P_{\mu}^{(1)}(k,-iD). $ Note also that
solutions to 
differential equation (\ref{4.4}) have to satisfy the boundary
condition $P_{\mu}^{(0)}(k,-iD) = -iD_{\mu} \equiv q_{\mu}.$
 The function $\Phi_{\mu\alpha}(D)$ in (\ref{4.4}) is deduced from (\ref{3.2}) and
 reads as
\begin{equation} \label{4.6}
  \Phi_{\mu\alpha}(D) =\eta_{\mu\alpha}(-A+f(B))+ia_{\mu}D_{\alpha}-(a^2-s)D_{\mu}D_\alpha\gamma_2.
\end{equation}
In all subsequent considerations we shall restrict ourselves to the case where $f(B)
= \sqrt{1-B}. $ Then we have $\gamma_2 = 0$ and consequently
Eq.(\ref{4.4}) reads
\begin{equation} \label{4.7}
 \frac{dP_{\mu}^{(t)}}{dt} ~ = ~ k_{\mu} \bigg[ aP^{(t)} +
 \sqrt{1+ (a^2 -s){(P^{(t)})}^2} ~ \bigg] -a_{\mu} kP^{(t)},
\end{equation}
where we have used an abbreviation $P_{\mu}^{(t)} \equiv P_{\mu}^{(t)}(k,-iD).$
The solution to differential equation (\ref{4.7}), which obeys the
required boundary conditions, looks as
\begin{eqnarray} \label{4.8}
 P_{\mu}^{(t)}(k,q) & = & q_{\mu} + \left( k_{\mu} Z^{-1}(q) -a_{\mu}
 (kq)  \right) \frac{\sinh(tW)}{W}  \\
        &+& \bigg[ \left(k_{\mu}(ak) - a_{\mu}k^2 \right) Z^{-1}(q)
     + a_{\mu}(ak) (kq) - sk_{\mu} (kq) \bigg] \frac{\cosh (tW)-1}{W^2}. \nonumber
\end{eqnarray}
In the above expression we have introduced the following abbreviations,
\begin{eqnarray} \label{4.9}
 W &=& \sqrt{{(ak)}^2 - s k^2}, \\
  Z^{-1} (q) & =& (aq) + \sqrt{1 + (a^2 - s)q^2}
\end{eqnarray}
and it is understood that quantities like $(kq)$ mean the scalar
product in a Minkowski space with signature
 $~ \eta_{\mu\nu} = diag (-1,1,\cdot \cdot \cdot, 1)$.
Now that we have $P_{\mu}^{(t)}(k,q),$ the required quantity 
$P_{\mu}(k,q)$ simply follows by setting $~t=1~$
and finaly we also get
\begin{eqnarray} \label{4.10}
  K_{\mu} (k)  = 
 \bigg[ k_{\mu} (ak)   -
  a_{\mu} k^2 \bigg] \frac{\cosh W -1}{W^2} 
    + k_{\mu} \frac{\sinh W}{W}.
\end{eqnarray}

Furthermore, we define the star product by the relation,
\begin{equation} \label{4.11}
  e^{ikX}~ \star ~ e^{iqX} ~ \equiv ~ e^{i K^{-1}(k) \x} ( e^{iqX})
 ~ = ~ e^{i {{\mathcal{D}}_{\mu} (k,q)}X^{\mu}}, 
\end{equation}
where
\begin{equation} \label{4.12}
   {\mathcal{D}}_{\mu} (k,q) ~ = ~ P_{\mu} (K^{-1}(k),q),
\end{equation}
with $K^{-1}(k)$ being the inverse of the transformation (\ref{4.10}).

We claim that the function $~{\mathcal{D}}_{\mu} (k,q)~$ is related to
   the coproduct $\triangle D_{\mu}$ for the translation
   generators. In particular, this relation is given by
\begin{equation} \label{4.12a}
  i{\mathcal{D}}_{\mu} (-iD \otimes 1, 1 \otimes (-iD)) = \triangle
  D_{\mu}. 
\end{equation} 
   This can be seen as follows \cite{Meljanac:2006ui,KresicJuric:2007nh,Meljanac:2007xb}.
 We start with the general definition
   of the star product 
$$
 f~ \star ~ g = m_{\star} (f \otimes g),
$$
where $m_{\star}$ denotes deformed multiplication in the commutative
algebra of smooth functions. If we take a derivative of both sides in
the above definition of the star product, we get
$$
 D_{\mu} (f \star g) = m_{\star} (\triangle D_{\mu} (f \otimes g)).
$$
By applying this to (\ref{4.11}), we find (\ref{4.12a}).

  Thus, the function $~{\mathcal{D}}_{\mu} (k,q)~$ determines the deformed
   Leibniz rule and the corresponding coproduct $\triangle D_{\mu}.$
   It also gives a generalized momentum addition rule.
  However, in the general case of
   deformation, when both parameters $a_{\mu}$ and $s$ are different
   from zero, it is quite a difficuilt task to obtain a closed form
   for $\triangle D_{\mu},$ so we give it in a form of a series
   expansion up to second order in the deformation parameter $a,$
\begin{eqnarray} \label{4.16}
  \triangle D_{\mu} &=&  D_{\mu}\otimes \mathbf{1} +\mathbf{1}\otimes
 D_{\mu} \nonumber \\
 & -& iD_{\mu} \otimes aD +
 ia_{\mu} D_{\alpha} \otimes D^{\alpha}
 -\frac{1}{2}(a^2-s)D_{\mu}\otimes D^2 \\
 & -& a_{\mu} (aD)D_{\alpha} \otimes D^{\alpha}
   +\frac{1}{2} a_{\mu} D^2 \otimes aD
   +\frac{1}{2} s D_{\mu} D_{\alpha} \otimes D^{\alpha} + {\mathcal{O}}(a^3). \nonumber
\end{eqnarray} 

Now that we have a coproduct, it is a straightforward procedure
 \cite{Meljanac:2006ui,Meljanac:2007xb}
to construct a star product between arbitrary
two functions $f$ and $g$ of commuting coordinates, generalizing in this way relation (\ref{4.11}) that
holds for plane waves. Thus, the general result for the star product, valid for the
NC space (\ref{2.1}), has the form
\begin{equation} \label{4.17}
 (f~ \star ~ g)(X) = \lim_{\substack{Y \rightarrow X  \\ Z \rightarrow X }}
 e^{X_{\alpha} [ i{\mathcal{D}}^{\alpha}(-iD_{Y},
 -iD_{Z}) - D_{Y}^{\alpha} -D_{Z}^{\alpha} ]} f(Y)g(Z).
\end{equation} 
Although star product is a binary operation acting on the algebra of functions
defined on the ordinary commutative space, it encodes features
that reflect noncommutative nature of space (\ref{2.1}). Note also
that in the case when  $s$ is different from zero, the star product (\ref{4.17})
is nonassociative.

The general results
obtained so far can be specialized to some particular cases, of which
Snyder space and $\kappa$-Minkowski space are particularly interesting.
   
 For $a^2=0,$ we have a Snyder type of noncommutativity,  
\begin{equation}
[\x_{\mu},\x_{\nu}]=s M_{\mu \nu}.
\end{equation}
In this situation, our realization (\ref{3.2}) reduces precisely to
that obtained in \cite{Battisti:2008xy}. For a special choice when
 $f(B) =1,$ we have the realization
\begin{equation}
\x_{\mu}=X_{\mu}-s(XD)D_\mu,
\end{equation}
which is the  case that was also considered in \cite{Licht:2005rm}.
In other interesting situation when $f(B)=\sqrt{1-B},$ the general result
(\ref{3.2}) reduces to
\begin{equation}
\x_{\mu}=X_{\mu} \sqrt{1+sD^2}.
\end{equation}
This choice of $f(B)$ is the one for which most of our results, through
all over
the paper, are obtained and which is the main subject of our
investigations. It is also considered by Maggiore \cite{Maggiore:1993rv,Maggiore1}.
 For this case when $f(B)=\sqrt{1-B},$ the exact
result for the coproduct (\ref{4.12}) can be obtained and it is given by
\begin{equation} \label{snydercoproduct}
\triangle D_{\mu} = D_{\mu}\otimes Z^{-1}+\mathbf{1}\otimes D_{\mu} 
 + s D_{\mu} D_{\alpha} \frac{1}{Z^{-1} +1} \otimes D^{\alpha},
\end{equation}
where
\begin{equation} 
Z^{-1} = \sqrt{1+sD^2}.
\end{equation}

Another interesting case is 
 when the parameter $\; s \;$ is equal to zero. This corresponds
 to $\kappa$-deformed space investigated in 
\cite{Lukierski:1991pn,Lukierski:1993wx,KowalskiGlikman:2002we,KowalskiGlikman:2002jr} and
\cite{KresicJuric:2007nh,Meljanac:2007xb}.
The general
 form (\ref{3.2}) for the realizations in the case
of $\kappa$-deformed space reduces to
\begin{align}
\x_{\mu} = X_{\mu} \left(-A+\sqrt{1-B}\right)+i(aX)\, D_{\mu},
\end{align}
with $B= a^2 D^2, $ and the coproduct takes on the form
\begin{equation}\label{coproductD}
\triangle D_{\mu} = D_{\mu}\otimes Z^{-1}+\mathbf{1}\otimes
D_{\mu}+ia_{\mu} (D_{\alpha} Z)\otimes
D^{\alpha}-\frac{ia_{\mu}}{2} \square\, Z\otimes iaD,
\end{equation}
where the generalized shift operator (\ref{16}) is here specialized to
\begin{align}
Z^{-1}= -iaD+\sqrt{1- a^2 D^2}.
\end{align}
The result (\ref{coproductD}) is exact and is also written in a closed form, as is the
coproduct (\ref{snydercoproduct}) for the case of Snyder space.

\section{Conclusion}

   In this paper we have investigated a Lie-algebraic type of
    deformations of Minkowski space and analyzed
    the impact these
    deformations have on the modification of  coalgebraic structure of
    the symmetry algebra underlying Minkowski space. By finding a coproduct, we were  able
    to see how coalgebra, which encodes the deformation of
    Minkowski space, modifies and to which extent 
   the Leibniz rule is deformed with respect to ordinary Leibniz rule. Since the
   coproduct is related to the star product, we were also able to write
    how star product looks like on NC spaces characterized by the
    general class of deformations of type (\ref{2.1}).
     We have also found many different  classes of realizations of NC space (\ref{2.1})
   and specialized obtained results to some specific cases of
    particular interest.

    The deformations that we have considered  
     are characterized by the common feature that the algebraic sector
     of the quantum (Hopf) algebra, which
     is described by the Poincar\'{e} algebra, is  undeformed, while, on the
     other hand, the corresponding coalgebraic sector is 
     affected by deformations.

      There is a vast variety of possible physical applications which
      could be expected to originate from the modified geometry
      at the Planck scale, which in turn reflects itself in a noncommutative
      nature of the configuration space. Which type of
      noncommutativity is inherent to configuration space is still not clear,
   but it is reasonable to expect that more wider is the class of
      noncommutativity taken into account,
   more likely is that it 
      will reflect true properties of geometry and relevant features at Planck scale.
  In particular, NC space considered in this paper is an interpolation of two types
      of noncommutativity, $\kappa$-Minkowski and Snyder, and as such
      is more likely  to reflect geometry at small distances
      then are each  of these spaces alone, at least, 
      it includes all features of both of these two types of
      noncommutativity, at the same time. As was already done
    for $\kappa$-type noncommutativity, it would be as well interesting to
      investigate the effects of combined $\kappa$-Snyder
      noncommutativity on dispersion relations
 \cite{AmelinoCamelia:2009pg,Magueijo1,Magueijo:2001cr}, black hole horizons \cite{Kim:2007nx},
       Casimir energy \cite{Kim:2007mb} and violation of CP
      symmetry, the problem that is considered in \cite{Glinka:2009ky} in the context of
      Snyder-type of noncommutativity. 

     Work that still remains to be done includes an elaboration and
   development of methods for physical theories on NC space considered
   here,
  particularly, the calculation of coproduct for the Lorentz
   generators, $\triangle M_{\mu\nu},$ $S$-antipode,
   differential forms \cite{Meljanac:2008pn,Kim:2008mp,Kim:1},
 Drinfeld twist \cite{Borowiec:2004xj,Borowiec:1,borowiec,Bu:2006dm,Govindarajan:2008qa},
   twisted flip operator \cite{Govindarajan:2008qa,Young:2007ag,Young:1,gov}
 and $ R$-matrix \cite{Young:2008zm,gov}.
    We shall address these issues in the forthcoming papers, together
   with a number of physical applications, such as the field theory
   for scalar fields \cite{Daszkiewicz:2007az,Daszkiewicz:1,Daszkiewicz:2,Freidel:2006gc}
 and its twisted statistics properties, as a
   natural continuation of our investigations put forward in previous
   papers \cite{Govindarajan:2008qa,gov}.

\section*{Acknowledgments}

  We thank A. Borowiec and Z. \v{S}koda for useful comments. 
     This work was supported by the Ministry of Science and Technology
    of the Republic of Croatia under contract No. 098-0000000-2865.


\begin{thebibliography}{0}
\bibitem{Doplicher:1994zv}
  S.~Doplicher, K.~Fredenhagen and J.~E.~Roberts,
  Phys.\ Lett.\  B {\bf 331} (1994) 39.
\bibitem{Doplicher:1994tu}
  S.~Doplicher, K.~Fredenhagen and J.~E.~Roberts,
  Commun.\ Math.\ Phys.\  {\bf 172} (1995) 187.

\bibitem{AmelinoCamelia:2000zs} G.~Amelino-Camelia and T.~Piran,
  Phys.\ Rev.\  D {\bf 64} (2001) 036005.

\bibitem{Seiberg:1999vs}
  N.~Seiberg and E.~Witten,
  JHEP {\bf 9909} (1999) 032.

\bibitem{AmelinoCamelia:2000mn}
  G.~Amelino-Camelia,
  Int.\ J.\ Mod.\ Phys.\  D {\bf 11} (2002) 35.

\bibitem{AmelinoCamelia:2000ge}
  G.~Amelino-Camelia,
  Phys.\ Lett.\  B {\bf 510} (2001) 255.

\bibitem{Lukierski:1991pn}
  J.~Lukierski, H.~Ruegg, A.~Nowicki and V.~N.~Tolstoi,
  Phys.\ Lett.\  B {\bf 264} (1991) 331.

\bibitem{Lukierski:1993wx}
  J.~Lukierski, H.~Ruegg and W.~J.~Zakrzewski,
  Annals Phys.\  {\bf 243} (1995) 90.

\bibitem{KowalskiGlikman:2002we}
  J.~Kowalski-Glikman and S.~Nowak,
  Phys.\ Lett.\  B {\bf 539} (2002) 126.

\bibitem{KowalskiGlikman:2002jr}
  J.~Kowalski-Glikman and S.~Nowak,
  Int.\ J.\ Mod.\ Phys.\  D {\bf 12} (2003) 299.

\bibitem{snyder} H. S. Snyder, Phys. Rev. {\bf 71} (1947) 38.

\bibitem{Ghosh:2006cb}
  S.~Ghosh,
  Phys.\ Rev.\  D {\bf 74} (2006) 084019.
\bibitem{Ghosh1}
  S.~Ghosh,
  Phys. Lett. B {\bf 648} (2007) 262. 

\bibitem{Chatterjee:2008bp}
  C.~Chatterjee and S.~Gangopadhyay,
  Europhys.\ Lett.\  {\bf 83} (2008) 21002.

\bibitem{Meljanac:2006ui}
  S.~Meljanac and M.~Stojic,
  Eur.\ Phys.\ J.\  C {\bf 47} (2006) 531, arXiv:hep-th/0605133.

\bibitem{KresicJuric:2007nh}
  S.~Kresic-Juric, S.~Meljanac and M.~Stojic,
  Eur.\ Phys.\ J.\  C {\bf 51} (2007) 229, arXiv:hep-th/0702215.

\bibitem{Meljanac:2007xb}
  S.~Meljanac, A.~Samsarov, M.~Stojic and K.~S.~Gupta,
  Eur.\ Phys.\ J.\  C {\bf 53} (2008) 295, arXiv:0705.2471 [hep-th].

\bibitem{Jonke:2002kb}
  L.~Jonke and S.~Meljanac,
  Eur.\ Phys.\ J.\  C {\bf 29} (2003) 433, arXiv:hep-th/0210042.
\bibitem{Jonke1}
  I.~Dadic, L.~Jonke and S.~Meljanac,
  Acta Phys.\ Slov.\  {\bf 55} (2005) 149, arXiv:hep-th/0301066.

\bibitem{Durov:2006iv}
  N.~Durov, S.~Meljanac, A.~Samsarov and Z.~Skoda,
  J.\ Algebra {\bf 309} (2007) 318, arXiv:math/0604096 [math.RT].

\bibitem{Meljanac:2008ud}
  S.~Meljanac and S.~Kresic-Juric,
  J.\ Phys.\ A  {\bf 41} (2008) 235203, arXiv:0804.3072 [hep-th].

\bibitem{Meljanac:2008pn}
  S.~Meljanac and S.~Kresic-Juric,
  J. Phys. A {\bf 42} (2009) 365204, arXiv:0812.4571 [hep-th].

\bibitem{Doresic:1994zz}
S. Meljanac, M. Milekovi{\'c} and  S. Pallua,
 Phys. Lett. B {\bf 328} (1994) 55, arXiv:hep-th/9404039.
\bibitem{palmil1}
 S. Meljanac and M. Milekovi{\'c},
 Int. J. Mod. Phys. A {\bf 11} (1996) 1391.
\bibitem{palmil2}
 S. Meljanac, M. Milekovi{\'c} and M. Stoji{\'c},
 Eur. Phys. J. C {\bf 24} (2002) 331, arXiv:math-ph/0201061.

\bibitem{bardek} V. Bardek and S. Meljanac,
 Eur. Phys. J. C {\bf 17} (2000) 539, arXiv:hep-th/0009099.
\bibitem{bardek1}
V. Bardek, L. Jonke, S. Meljanac and  M. Milekovi{\'c},
 Phys. Lett. B {\bf 531} (2002) 311, arXiv:hep-th/0107053v5.
\bibitem{bardek2}
 L. Jonke and S. Meljanac,
 Phys. Lett. B {\bf 526} (2002) 149, arXiv:hep-th/0106135.

\bibitem{Battisti:2008xy}
  M.~V.~Battisti and S.~Meljanac,
  Phys. Rev. D {\bf 79} (2009) 067505, arXiv:0812.3755 [hep-th].

\bibitem{Guo:2008qp}
  H.~Y.~Guo, C.~G.~Huang and H.~T.~Wu,
  Phys.\ Lett.\  B {\bf 663} (2008) 270.

\bibitem{Romero:2004er}
  J.~M.~Romero and A.~Zamora,
  Phys.\ Rev.\  D {\bf 70} (2004) 105006.

\bibitem{Banerjee:2006wf}
  R.~Banerjee, S.~Kulkarni and S.~Samanta,
  JHEP {\bf 0605} (2006) 077.

\bibitem{Glinka:2008tr}
  L.~A.~Glinka,
  Apeiron {\bf 16} (2009) 147.

\bibitem{Licht:2005rm}
  A.~L.~Licht,
  arXiv:hep-th/0512134.

\bibitem{Maggiore:1993rv}
  M.~Maggiore,
  Phys.\ Lett.\  B {\bf 304} (1993) 65.
\bibitem{Maggiore1}
  M.~Maggiore,
  Phys.\ Rev.\  D {\bf 49} (1994) 5182.

\bibitem{AmelinoCamelia:2009pg}
  G.~Amelino-Camelia and L.~Smolin,
  arXiv:0906.3731 [astro-ph.HE].

\bibitem{Magueijo:2001cr}
  J.~Magueijo and L.~Smolin,
  Phys.\ Rev.\ Lett.\  {\bf 88} (2002) 190403.
\bibitem{Magueijo1}
  J.~Magueijo and L.~Smolin,
  Phys.\ Rev.\  D {\bf 67} (2003) 044017.

\bibitem{Kim:2007nx}
  H.~C.~Kim, M.~I.~Park, C.~Rim and J.~H.~Yee,
  JHEP {\bf 0810} (2008) 060.

\bibitem{Kim:2007mb}
  H.~C.~Kim, C.~Rim and J.~H.~Yee,
  arXiv:0710.5633 [hep-th].

\bibitem{Glinka:2009ky}
  L.~A.~Glinka,
  arXiv:0902.4811 [hep-ph].  

\bibitem{Kim:2008mp}
  H.~C.~Kim, Y.~Lee, C.~Rim and J.~H.~Yee,
  Phys.\ Lett.\  B {\bf 671} (2009) 398.
\bibitem{Kim:1}
  J.~G.~Bu, J.~H.~Yee and H.~C.~Kim,
arXiv:0903.0040 [hep-th].

\bibitem{Borowiec:2004xj}
  A.~Borowiec, J.~Lukierski and V.~N.~Tolstoy,
  Eur.\ Phys.\ J.\  C {\bf 44} (2005) 139.
\bibitem{Borowiec:1}
  A.~Borowiec, J.~Lukierski and V.~N.~Tolstoy,
  Eur.\ Phys.\ J.\  C {\bf 48} (2006) 633.
\bibitem{borowiec}
 A. Borowiec and A. Pachol, 
Phys.Rev.D {\bf 79} (2009) 045012. 

\bibitem{Bu:2006dm}
  J.~G.~Bu, H.~C.~Kim, Y.~Lee, C.~H.~Vac and J.~H.~Yee,
  Phys.\ Lett.\  B {\bf 665} (2008) 95.

\bibitem{Govindarajan:2008qa}
  T.~R.~Govindarajan, K.~S.~Gupta, E.~Harikumar, S.~Meljanac and D.~Meljanac,
  Phys.\ Rev.\  D {\bf 77} (2008) 105010, arXiv:0802.1576 [hep-th].

\bibitem{Young:2007ag}
  C.~A.~S.~Young and R.~Zegers,
  Nucl.\ Phys.\  B {\bf 797} (2008) 537.
\bibitem{Young:1}
  C.~A.~S.~Young and R.~Zegers,
  Nucl.\ Phys.\  B {\bf 804} (2008) 342.

\bibitem{gov}
  T.~R.~Govindarajan, K.~S.~Gupta, E.~Harikumar, S.~Meljanac and D.~Meljanac,
  Phys. Rev. D {\bf 80} (2009) 025014, arXiv:0903.2355 [hep-th].

\bibitem{Young:2008zm}
  C.~A.~S.~Young and R.~Zegers,
  Nucl.\ Phys.\  B {\bf 809} (2009) 439.

\bibitem{Daszkiewicz:2007az}
  M.~Daszkiewicz, J.~Lukierski and M.~Woronowicz,
  Mod.\ Phys.\ Lett.\  A {\bf 23} (2008) 653.
\bibitem{Daszkiewicz:1}
  M.~Daszkiewicz, J.~Lukierski and M.~Woronowicz,
  Phys.\ Rev.\  D {\bf 77} (2008) 105007.
\bibitem{Daszkiewicz:2}
  M.~Daszkiewicz, J.~Lukierski and M.~Woronowicz,
  J.\ Phys.\ A  {\bf 42} (2009) 355201.

\bibitem{Freidel:2006gc}
  L.~Freidel, J.~Kowalski-Glikman and S.~Nowak,
  Phys.\ Lett.\  B {\bf 648} (2007) 70.

\end{thebibliography}
\end{document}